\documentclass{emulateapj}
\usepackage{float}
\usepackage{amsmath}
\usepackage{longtable}
\usepackage{color}

\begin{document}

\title{Stellar Absorption Line Analysis of Local Star-Forming Galaxies: The Relation Between Stellar Mass, Metallicity, Dust Attenuation and Star Formation Rate}

\author{H. Jabran Zahid$^{1}$\footnote{email: zahid@cfa.harvard.edu}, Rolf-Peter Kudritzki$^{2}$, Charlie Conroy$^{3}$, Brett Andrews$^{4}$ \& I-Ting Ho$^{2,5}$}
\affil{$^{1}$Smithsonian Astrophysical Observatory, Harvard-Smithsonian Center for Astrophysics - 60 Garden Street, Cambridge, MA 02138, USA}
\affil{$^{2}$University of Hawaii at Manoa, Institute for Astronomy - 2680 Woodlawn Drive, Honolulu,  HI 96822, USA}
\affil{$^{3}$Department of Astronomy, Harvard University, Cambridge, MA, 02138, USA}
\affil{$^{4}$PITT PACC, Department of Physics and Astronomy, University of Pittsburgh, 3941 O'Hara St., Pittsburgh, PA 15260, USA}
\affil{$^{5}$Galaxies and Cosmology Department - Max-Planck Institute for Astronomy - K\"{o}nigstuhl 17, 69117 Heidelberg, Germany}

\date{}                                           
\begin{abstract}

We analyze the optical continuum of star-forming galaxies in SDSS by fitting stacked spectra with stellar population synthesis models to investigate the relation between stellar mass, stellar metallicity, dust attenuation and star formation rate. We fit models calculated with star formation and chemical evolution histories that are derived empirically from multi-epoch observations of the stellar mass---star formation rate and the stellar mass---gas-phase metallicity relations, respectively. We also fit linear combinations of single burst models with a range of metallicities and ages. Star formation and chemical evolution histories are unconstrained for these models. The stellar mass---stellar metallicity relations obtained from the two methods agree with the relation measured from individual supergiant stars in nearby galaxies. These relations are also consistent with the relation obtained from emission line analysis of gas-phase metallicity after accounting for systematic offsets in the gas-phase-metallicity. We measure dust attenuation of the stellar continuum and show that its dependence on stellar mass and star formation rate is consistent with previously reported results derived from nebular emission lines. However, stellar continuum attenuation is smaller than nebular emission line attenuation. The continuum-to-nebular attenuation ratio depends on stellar mass and is smaller in more massive galaxies. Our consistent analysis of stellar continuum and nebular emission lines paves the way for a comprehensive investigation of stellar metallicities of star-forming and quiescent galaxies.

\end{abstract}

\section{Introduction}

Measurements of heavy elements and dust provide important constraints for understanding the formation and evolution of star-forming galaxies. Gas from the intergalactic medium flows into dark matter halos fueling star formation in galaxies. Stars are sustained by fusion of lighter elements into heavier elements. These heavy elements are recycled into the interstellar medium (ISM) by stellar mass loss processes. Some fraction of these heavy elements may be expelled from the ISM by galaxy scale outflows. With each generation of star formation, heavy elements and dust---which forms out of heavy elements---accumulate in galaxies. Thus, the heavy element and dust content of galaxies depends on star formation and gas flows. These are key physical processes governing galaxy formation and evolution.

Oxygen is the most abundant heavy element in the universe. The amount of oxygen relative to hydrogen in the gas-phase is an important metric of chemical evolution. The gas-phase metallicity of star-forming galaxies can be measured from strong emission lines observed in rest-frame optical spectra \citep{Searle1972}. \cite{Lequeux1979} first showed that gas-phase metallicity of star-forming galaxies scales with stellar mass---the so-called mass---metallicity ($MZ$) relation. From analysis of $\sim50,000$ star-forming galaxies in the Sloan Digital Sky Survey (SDSS), it is now well established that there is a tight $MZ$ relation for galaxies in the local universe \citep[$\sim0.1$ dex scatter;][]{Tremonti2004}. The $MZ$ relation is a power-law that flattens or saturates at large stellar masses \citep{Tremonti2004, Moustakas2011, Andrews2013, Zahid2013b, Wu2017}. 

An $MZ$ relation is observed for star-forming galaxies out to $z\sim3$ \citep[e.g.,][]{Savaglio2005, Erb2006b, Cowie2008, Maiolino2008, Mannucci2009, Lamareille2009, Perez-Montero2009a, Zahid2011a, Yabe2012, Zahid2013b, Zahid2014c, Wuyts2014, Maier2015, Salim2015, Sanders2015, Ly2016, Kashino2017}. These observations demonstrate that at a fixed stellar mass, metallicities of galaxies increase with cosmic time.  \citet{Zahid2013b} show that the shape and overall normalization of the $MZ$ relation is independent of redshift at $z\lesssim2$; the evolution of the $MZ$ relation is quantified solely by evolution in the stellar mass where metallicities of galaxies saturate, i.e. the stellar mass where the $MZ$ relation flattens. 

The scatter of the $MZ$ relation is correlated with other galaxy properties. A relation between stellar mass, metallicity and SFR is observed in local \citep[SFR][]{Ellison2008, Mannucci2010, Lara-Lopez2010, Yates2012, Andrews2013, Salim2014} and high-redshift galaxies \citep{Zahid2014c, Yabe2014, Troncoso2014, Maier2015, Salim2015}. At a fixed stellar mass, galaxies with high SFRs tend to have lower metallicities and vice versa. Both metallicity and SFR depend on the amount of gas. Thus, the anti-correlation between metallicity and SFR is likely due to variations in gas content \citep{Hughes2013, Bothwell2013, Bothwell2016}.

The shape, evolution and correlated scatter of the $MZ$ relation provide important observational constraints for models of galactic chemical evolution \citep[e.g.,][]{Ellison2008, Finlator2008, Mannucci2010, Dave2012, Dayal2013, Lilly2013, Zahid2013b, Zahid2014b, Ascasibar2015, Andrews2017}. Based on these observational constraints and the equations of galactic chemical evolution, \citet[Z14 hereafter]{Zahid2014b} develop an analytical model whereby the $MZ$ relation originates from a universal relation between metallicity and stellar-to-gas-mass ratio. In their model, star formation is fueled by inflows and metallicity increases as galaxies build stellar mass. However, the metallicity saturates at the point when the mass of metals produced and returned to the ISM by massive stars is equal to the mass of metals forever sequestered within low mass stars. 

Stars form from gas composing the ISM and thus a relation between stellar mass and stellar metallicity is expected. Indeed, several studies have reported such a relation between the stellar mass and the metallicity of individual stars and/or the integrated stellar population \citep{Gallazzi2005, Panter2008, Sommariva2012, Kirby2013, Gonzalez-Delgado2014, Gonzalez-Delgado2015, Kudritzki2016, Bresolin2016}. The stellar $MZ$ relation is qualitatively similar to the gas-phase $MZ$ relation. 

The gas-phase metallicity depends on SFR which is likely due to variations in gas content. If these variations occur on sufficiently long timescales, stellar metallicity determined from the integrated stellar population should also depend on SFR. The dependence of the stellar $MZ$ relation on the SFR will be explored in this work.

Dust forms from heavy elements and a correlation between dust and metallicity is observed in the local universe \citep{Heckman1998, Boissier2004, Asari2007, Garn2010b, Xiao2012, Zahid2012b} and at high redshifts \citep{Reddy2010, Zahid2014c}. At a fixed stellar mass, dust attenuation measured from the Balmer decrement depends on stellar mass and SFR \citep{Zahid2013a}. At stellar masses below $\lesssim10^{10} M_\odot$ dust attenuation is anti-correlated with SFR; at stellar masses $\gtrsim 10^{10} M_\odot$ dust attenuation and SFR are positively correlated. \citet{Yates2012} report similar trends for the relation between stellar mass, metallicity and SFR \citep[see also][]{Zahid2013a}. Here we report dust attenuation measurements derived from analysis of the stellar continuum. Dust attenuation determined from the continuum is independent of the emission line properties and thus provides an alternative means to investigate the relation between stellar mass, dust attenuation and SFR.

Studies of local galaxies report that nebular lines are more attenuated than the continuum \citep[e.g.,][]{Calzetti1994, Mayya1996, Charlot2000}. The canonical ratio of continuum-to-nebular attenuation is 0.44 \citep{Calzetti1997b}. However, this ratio may be a function of galaxy properties \citep[e.g.,][]{Wuyts2013, Koyama2015}. We examine the ratio of continuum-to-nebular attenuation as a function of galaxy properties.

The metal and dust content of star-forming galaxies is typically studied via emission line analysis of optical spectra. However, metallicities measured from strong emission lines are subject to large systematic uncertainties, which are not yet fully understood \citep{Kewley2008}. Therefore, we develop an alternative approach to analyze spectra and revisit relations between stellar mass, metallicity, dust content and SFR for star-forming galaxies in the SDSS. We derive metallicity and dust attenuation by analyzing metal absorption lines and the continuum shape of the integrated stellar population. Our analysis is based on fitting stacked spectra with stellar population synthesis models. This approach is completely independent of emission line analyses and thus provides a robust test of previously reported results.

In Section 2 and 3 we describe the observational data and methods, respectively. We present the $MZ$ relation in Section 4 and examine the relation between stellar mass and dust attenuation in Section 5. We investigate these quantities as a function of SFR in Section 6 and discuss our results in Section 7. We conclude in Section 8. We adopt $12+\mathrm{log}(O/H) = 8.69$ and $Z = 0.0142$ as the solar metallicity \citep{Asplund2009} and a fiducial cosmology $(H_0, \Omega_m, \Omega_\lambda)$ = (70 km s$^{-1}$, 0.3, 0.7). 

\section{{Observations}}

\subsection{SDSS data and Sample Selection}

We analyze the Main Galaxy Sample of the Sloan Digital Sky Survey \citep{Abazajian2009, Alam2015}. The parent sample consists of $\sim900,000$ galaxies observed over 10,000 deg$^2$ down to a limiting magnitude of $r < 17.8$ at $0 \lesssim z \lesssim 0.3$. The observed spectral range is $3800 - 9200 \mathrm{\AA}$ with a nominal spectral resolution of $R\sim1500$ at $5000\mathrm{\AA}$ \citep{Smee2013}. We adopt total stellar masses, total star formation rates and emission line fluxes measured by the MPA/JHU group\footnote{http://wwwmpa.mpa-garching.mpg.de/SDSS/DR7/} \citep{Kauffmann2003a, Tremonti2004,Brinchmann2004, Salim2007}.

We select galaxies following the criteria described in \citet{Andrews2013}. The sample is restricted to $0.027<z<0.25$ to ensure rest-frame wavelength coverage of the $[OII]\lambda3727$ and $[OII]\lambda{7330}$ emission lines. We remove galaxies classified as active galactic nuclei (AGN) using the \citet{Kauffmann2003b} classification based on the \citet{Baldwin1981} diagram. The classification is based on a ratio of emission line fluxes. Star-forming galaxies are classified as having line ratios
\begin{equation}
\begin{split}
 \mathrm{log}([OIII]\lambda5007/H\beta) >  ~~~~~~~~~~~~~~~~~~~~~~~~~~~~ \\
\frac{0.61}{\mathrm{log}([NII]\lambda6584/H\alpha) - 0.05} + 1.3
\end{split}
\end{equation}
and
\begin{equation}
\mathrm{log}([NII]\lambda6584/H\alpha) < -0.05.
\end{equation}
We select galaxies with $H\beta$, $H \alpha$ and $[NII]\lambda6584$ emission lines observed with a signal-to-noise (S/N) ratio $\geq 5$. These selection criteria yield a sample of $\sim200,000$ star-forming galaxies.

\subsection{Stacking Procedure}
\begin{figure*}
\begin{center}
\includegraphics[width = 2 \columnwidth]{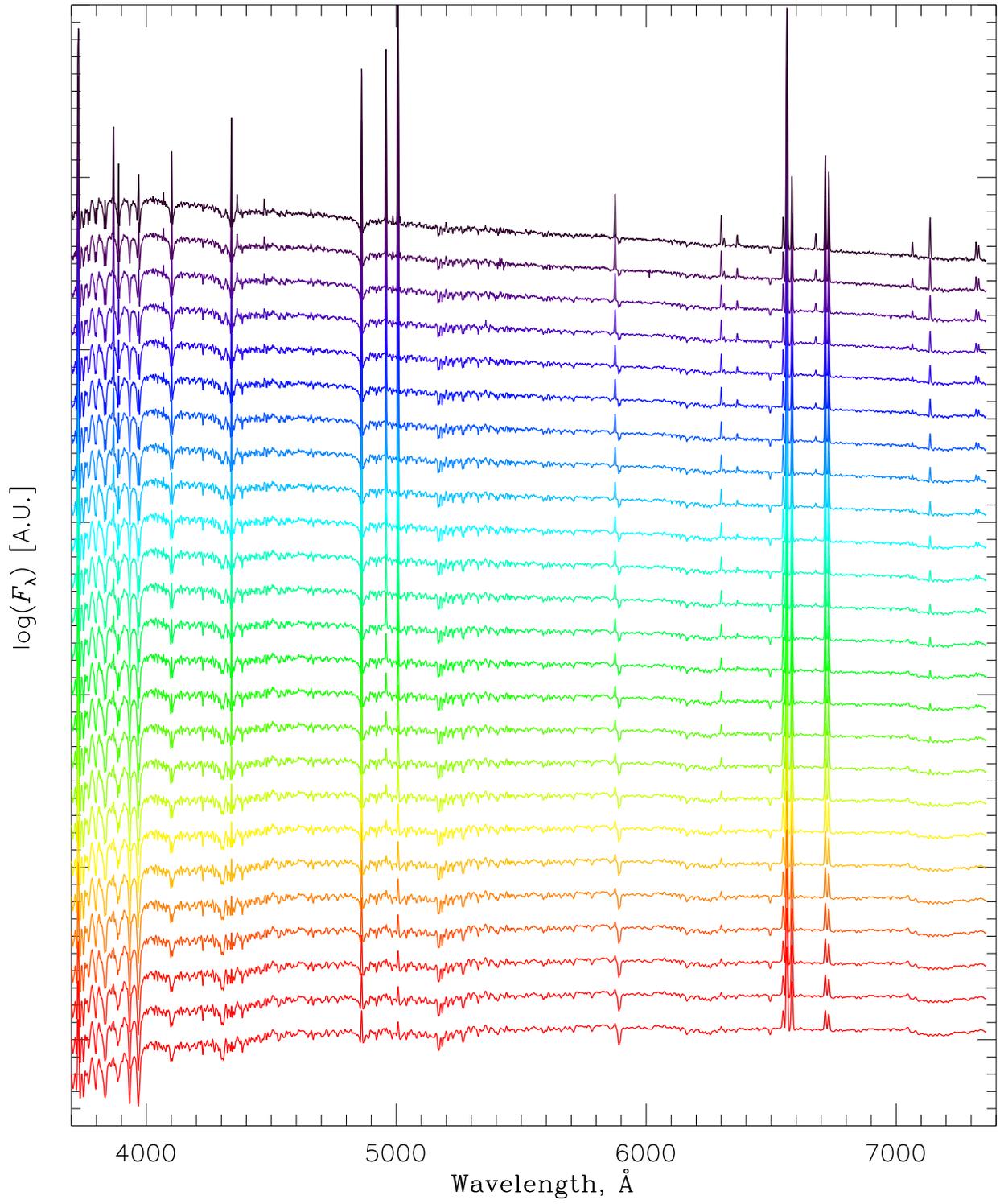}
\end{center}
\caption{SDSS spectra of star-forming galaxies stacked in 25 bins of stellar mass ranging from $10^{8.5} M_\odot$ (top) to $10^{11} M_\odot$ (bottom). Stellar mass ranges of stacked spectra are in Table \ref{tab:stack_prop}.}
\label{fig:montage}
\end{figure*}

 The techniques we develop in this work can be applied to individual galaxy spectra but the fitting procedure as it is currently implemented is too computationally expensive to apply to large samples. Thus, we stack spectra of galaxies in bins of stellar mass. This approach increases the S/N ratio of the spectra we analyze and requires fewer computations. The stacking procedure is described in detail by \citet{Andrews2013}. Here we highlight the salient aspects of the procedure and refer readers to AM13 for more details.

The data are stacked in 0.1 dex bins of stellar mass with stellar masses ranging between $10^{8.5} - 10^{11} M_\odot$. The spectra are corrected for galactic extinction using the maps of \citet{Schlegel1998} and are shifted to rest-frame wavelengths using the measured redshift. They are linearly interpolated between $3700-7360\mathrm{\AA}$ with a wavelength pixel resolution $\Delta \lambda = 1\mathrm{\AA}$ and are normalized to the mean flux between $4400 - 4450\mathrm{\AA}$. The spectra are coadded by taking the mean flux of all spectra in the stellar mass bin at each wavelength pixel element. Each spectrum is equally weighted in the average. To examine galaxy properties as a function of stellar mass and SFR, we sort spectra in each stellar mass bin into quintiles of SFR. Systematics related to the stacking procedure are discussed in AM13.

Figure \ref{fig:montage} shows a montage of the 25 stacked spectra analyzed in this study. The number of spectra in each stack and typical S/N ratio are given in Table \ref{tab:stack_prop}. We empirically determine the spectral resolution of each stacked spectrum from fitting strong emission lines and find that it increases with stellar mass. To avoid systematic effects caused by different spectral resolution, we convolve all spectra such that emission lines have the line width of the most massive galaxy stacked spectrum which is 330 km s$^{-1}$ (full width at half maximum).

{The 3" spectroscopic fiber aperture of SDSS is positioned on bright central region of galaxies and thus does not capture the total light. To test for aperture bias, we calculate the $g-$band covering fraction as the ratio of flux in the $g-$band fiber to the total $g-$band flux determined from the photometric magnitude. Thus, the $g-$band covering fraction is the fraction of the galaxy light in the $g-$band captured in the fiber. The median fiber aperture covering fraction is 0.25 for our sample and ranges between $0.05-1$.}

{ We test whether our results are sensitive to aperture bias. In each stellar mass bin, we sort the data into two equally populated bins of fiber aperture covering fraction; we have a large and small fiber aperture covering fraction stacked spectrum with typical covering fractions of 0.35 and 0.2, respectively. We analyze these data applying the procedure described in Section 3.2.2. The results of this exercise are consistent with results presented in this paper. Thus, we conclude that aperture bias is not a significant systematic effect \citep[][]{Choi2014, Peng2015} and the metallicity that we measure represents a global quantity averaged over the central regions of the galaxies. }

\begin{deluxetable}{cll}
\tablewidth{150pt}
\tablecaption{Stack Statistics}
\tablehead{\colhead{Stellar Mass Range} & \colhead{N$_S$} & \colhead{S/N} \\
                   log$(M_\ast/M_\odot)$ &}
\startdata

8.50$\,-\,$8.60    &        306  &    156  \\
8.60$\,-\,$8.70    &        558  &    197 \\
8.70$\,-\,$8.80    &        884  &    247 \\
8.80$\,-\,$8.90    &       1317  &    288 \\
8.90$\,-\,$9.00    &       1808  &    336 \\
9.00$\,-\,$9.10    &       2440  &    413 \\
9.10$\,-\,$9.20    &       3053  &    466\\
9.20$\,-\,$9.30    &       3931  &    528\\
9.30$\,-\,$9.40    &       4638  &    606\\
9.40$\,-\,$9.50    &       5592  &    695\\
9.50$\,-\,$9.60    &       6668  &    768\\
9.60$\,-\,$9.70    &       7852  &    888\\
9.70$\,-\,$9.80    &       8922  &    932\\
9.80$\,-\,$9.90    &      10155  &    989\\
9.90$\,-\,$10.0    &      11270  &    1067\\
10.0$\,-\,$10.1    &      12267  &    1140\\
10.1$\,-\,$10.2    &      13184  &    1209\\
10.2$\,-\,$10.3    &      13831  &    1275\\
10.3$\,-\,$10.4    &      14140 &    1332 \\
10.4$\,-\,$10.5    &      13958  &    1343\\
10.5$\,-\,$10.6    &      12990  &    1352\\
10.6$\,-\,$10.7    &      11652   &    1300 \\
10.7$\,-\,$10.8    &       9522  &    1223\\
10.8$\,-\,$10.9    &       7368  &    1082\\
10.9$\,-\,$11.0    &       5332  &    947\\
\enddata
\tablecomments{Columns 1 and 2 give the stellar mass range and number of spectra averaged, $N_S$, for each stack, respectively. Column 3 lists the median signal-to-noise per pixel between $4400-4450 \mathrm{\AA}$.}
\label{tab:stack_prop}
\end{deluxetable}

\section{Methods}

Our goal is to develop a robust technique to analyze the continuum of star-forming galaxies. We compare stellar metallicities determined from continuum analysis with those derived from standard techniques using ISM emission lines. We derive stellar metallicities by fitting stellar population synthesis models using two independent approaches. We empirically derive star formation and chemical evolution histories and use these as inputs to generate stellar population synthesis models which we fit to data. We also fit the data with a linear combination of models with single bursts of star formation which we sample at different metallicities and ages.

{The various approaches used to measure metallicity in this study are subject to different assumptions. We compare the results of the different methods to test these assumptions. Throughout this work we generically refer to the relation between metallicity and stellar mass as the $MZ$ relation. When specifically referring to the gas-phase and stellar mass-metallicity relations we denote them as $MZ_g$ and $MZ_\ast$, respectively.} 

{We derive gas-phase metallicities using standard technique of analyzing strong emission lines. The gas-phase metallicity is the abundance of oxygen relative to hydrogen. To convert the gas-phase abundance into total metallicity which we quote as $Z$, we assume a solar abundance pattern.}

{We derive evolutionary stellar population synthesis models which we fit to spectra stacked in bins of stellar mass. These models are based on empirically constrained star formation and chemical evolution histories that are derived from the average galaxy properties measured as a function of stellar mass. Thus, they are applicable to galaxy spectra stacked in bins of stellar mass. Models with prescribed histories have the advantage that the results can be easily interpreted because the star formation and chemical evolution histories are known. However, these models are only applicable when the spectrum being fit characterizes the average galaxy in a narrow range of stellar mass. These models are not relevant to spectra which are sorted by other galaxy properties. Thus, they cannot be applied to spectra sorted in quintiles of SFR. }

{The single burst model fitting procedure makes no assumptions about the star formation and chemical evolution history of galaxies. These models can be fit to galaxies spectra which are sorted by other galaxy properties such as SFR or to the spectrum of individual galaxies. The procedure assumes that the integrated stellar population is the sum of discrete star formation events with each event having a single age, metallicity and SFR. Our approach is to demonstrate the consistency of the two methods for determining stellar metallicities for data stacked only in bins of stellar mass. We then apply the single burst models to the stacked spectra sorted in quintiles of SFR.}

\subsection{Gas-Phase Oxygen Abundance}

The spectra in Figure \ref{fig:montage} show strong emission lines from ionized gas in star-forming regions. We measure the gas-phase oxygen abundance using line flux ratios corrected for dust attenuation. We correct emission line fluxes using the \citet{Cardelli1989} extinction law assuming case B recombination value for the H$\alpha$/H$\beta$ emission line ratio of 2.86 \citep{Hummer1987}. We determine the gas-phase oxygen abundance using the strong-line calibration of \citet[KK04 hereafter]{Kobulnicky2004}. The relevant emission line ratios are
\begin{equation}
R23 = \frac{[OII]\lambda3727 + [OIII]\lambda4959 + [OIII]\lambda5007}{H\beta}
\end{equation}
and
\begin{equation}
O32 = \frac{[OIII]\lambda4959 + [OIII]\lambda5007}{[OII]\lambda3727}.
\end{equation}
We assume that the $[OIII]\lambda5007$ to $[OIII]\lambda4959$ emission line ratio is 3:1 \citep{Osterbrock1989} and adopt 1.33 times $[OIII]\lambda5007$ line flux when summing the two lines. {The $R23$ line ratio has two metallicity branches. We adopt the lower metallicity branch when  log($[NII]/[OII]) < -1.2$ \citep{Kewley2008}}. Only a small fraction ($<1\%$) of galaxies are on the lower metallicity branch.

Metallicity diagnostics using strong emission lines are empirically and/or theoretically calibrated \citep[for review see][]{Kewley2008}. \citet{Kewley2008} show that different diagnostics applied to the same galaxies yield inconsistent measurements of metallicity. In particular, the absolute calibration varies by $\gtrsim 0.4$ dex. However, by comparing several diagnostics, they conclude that the $R23$ diagnostic used in this work is accurate in a relative sense. Thus, while the zero-point of the relation between stellar mass and gas-phase oxygen abundance is uncertain, the shape of the relation is more robust. 

\subsection{Stellar Population Synthesis Modeling and Fitting}

\begin{figure*}
\begin{center}
\includegraphics[width = 2 \columnwidth]{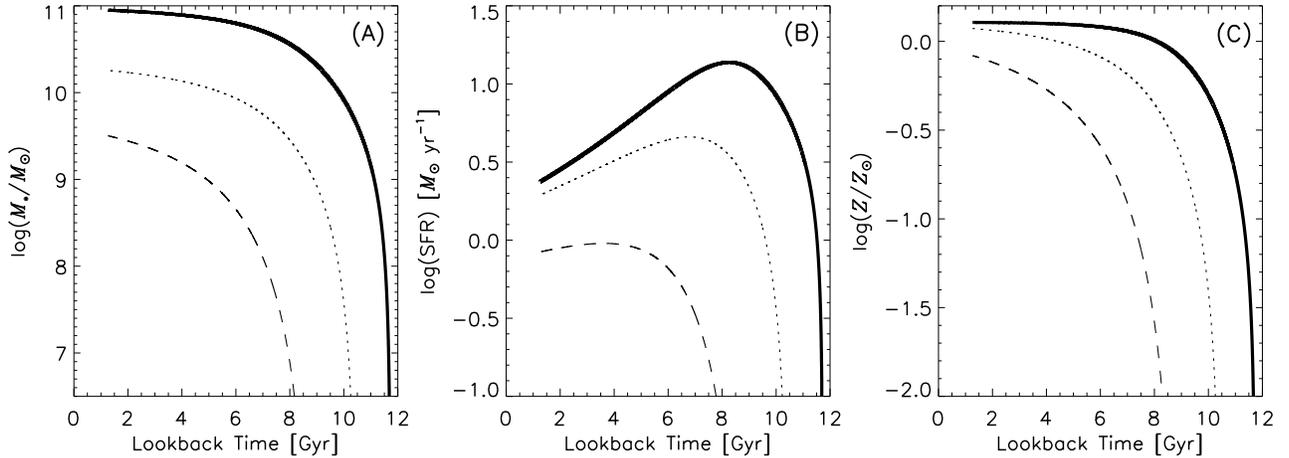}
\end{center}
\caption{(A) Stellar mass and (B) star formation history of galaxies that evolve along the mean relation between stellar mass and SFR observed at several epochs. (C) Metallicity history of the same model galaxies assuming they evolve along the mean $MZ_g$ relation measured at several epochs. The dashed, dotted and solid curves are examples of three model galaxies with stellar masses of $\sim10^{9.5}$, $10^{10.25}$ and $10^{11} M_\odot$ at $z\sim0$. The stellar mass and metallicity histories derived from these models are used as inputs to generate FSPS model spectra.}
\label{fig:z_sfh}
\end{figure*}

In addition to nebular emission lines, spectra in Figure \ref{fig:montage} show absorption lines in the continuum emitted by the integrated stellar population. These absorption lines encode information about stellar metallicity. \citet{Kudritzki2016} (and references therein) demonstrate that stellar absorption line spectroscopy is an accurate tool to determine metallicity of individual blue supergiant stars in nearby galaxies. Here we apply a similar spectral fitting technique except we use model spectra calculated with a stellar population synthesis approach rather than model spectra of individual stars.

Strong ISM emission and absorption lines as well as strong stellar Balmer lines contaminate the stacked stellar metal line spectra at certain wavelengths. When fitting stacked spectra with models, we avoid the near ultraviolet part of the spectrum which is dominated by a series of absorption lines (Calcium H,K and Balmer series) and the [OII]$\lambda3727$ emission line and limit our analysis to the spectral range of $4050 - 7300\mathrm{\AA}$. In this range we mask out a 30$\mathrm{\AA}$ region centered on each ISM absorption and emission line. However, for the Balmer lines we mask out 75$\mathrm{\AA}$ region to exclude the pressure broadened line profile wings. These masking windows are appropriate for the spectral resolution of the data.

We calculate model spectra using the Flexible Stellar Population Synthesis (FSPS; v3.0) code \citep{Conroy2009a, Conroy2010}. We adopt the \citet{Chabrier2003} initial mass function, the Medium-resolution Isaac Newton Telescope Library of Empirical Spectra (MILES) library \citep{Sanchez-Blazquez2006} and the Mesa Isochrones and Stellar Tracks \citep[MIST;][]{Dotter2016, Choi2016}. The intrinsic resolution of the model spectra is $2.5\mathrm{\AA}$ \citep{Falcon-Barroso2011, Beifiori2011} which we convolve to match the resolution of the stacked spectra (i.e. 330 km s$^{-1}$).

\subsubsection{{Metallicity from Look Back Models}}

We calculate ``look back'' models by inputting empirically constrained star formation and chemical evolution histories. Star formation history is not a free parameter, thus we do not need the observed shape of the spectrum of the integrated stellar population as a constraint for the analysis. We normalize stacked and model spectra by fitting a polynomial of degree 20. We divide the spectrum by the best-fit polynomial. Normalizing spectra to the continuum means that effects related to dust attenuation are not an issue for this part of the analysis.

\begin{figure*}
\begin{center}
\includegraphics[width = 2 \columnwidth]{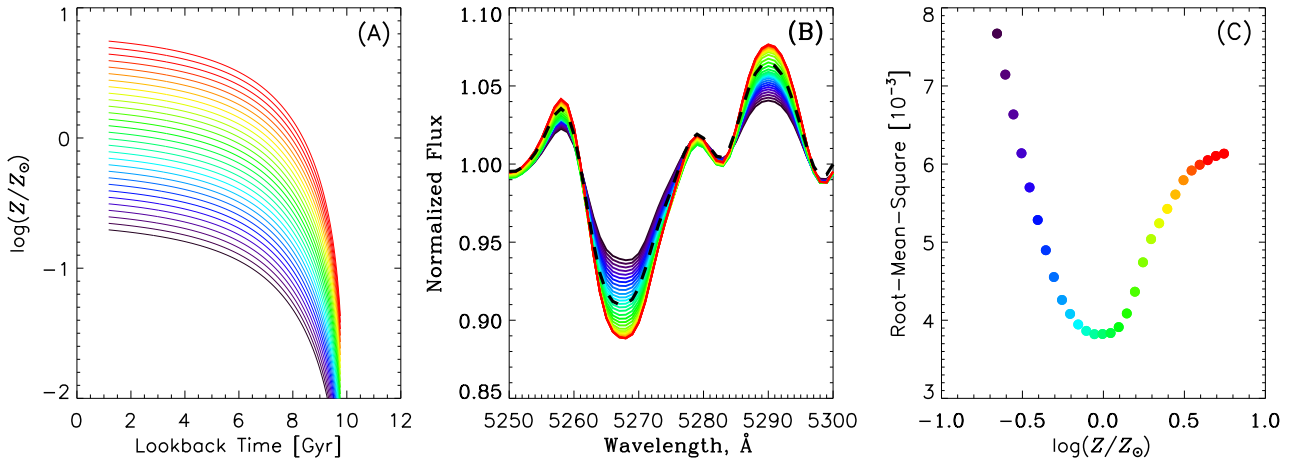}
\end{center}
\caption{Example of an FSPS look back model spectrum fit. (A) Chemical evolution history for evolution models with different offsets $\Delta Z_0$ of the zero-point $Z_0$ in Equation \ref{eq:mzfit}. Offsets range between $-0.75 < \Delta Z_0 < 0.75$ (for details, see text). (B) Example of model template spectra in one narrow spectral window calculated using FSPS and tabulated star formation and metallicity history based on the models shown in Figure \ref{fig:z_sfh}. Colors correspond to (A). The dashed line is the observed stacked spectrum for galaxies with $10^{10} \leq M_\ast /  M_\odot \leq 10^{10.1}$. (C) RMS difference between the stacked spectrum and model spectra plotted as a function of final metallicity. The best-fit metallicity is determined by fitting a parabola to points around the minimum RMS. The best-fit metallicity represents the final metallicity of a galaxy evolving along the $MZ_g$ relation and is interpreted as the metallicity of the young stellar population.}
\label{fig:chi2_template}
\end{figure*}

\begin{figure}
\begin{center}
\includegraphics[width = \columnwidth]{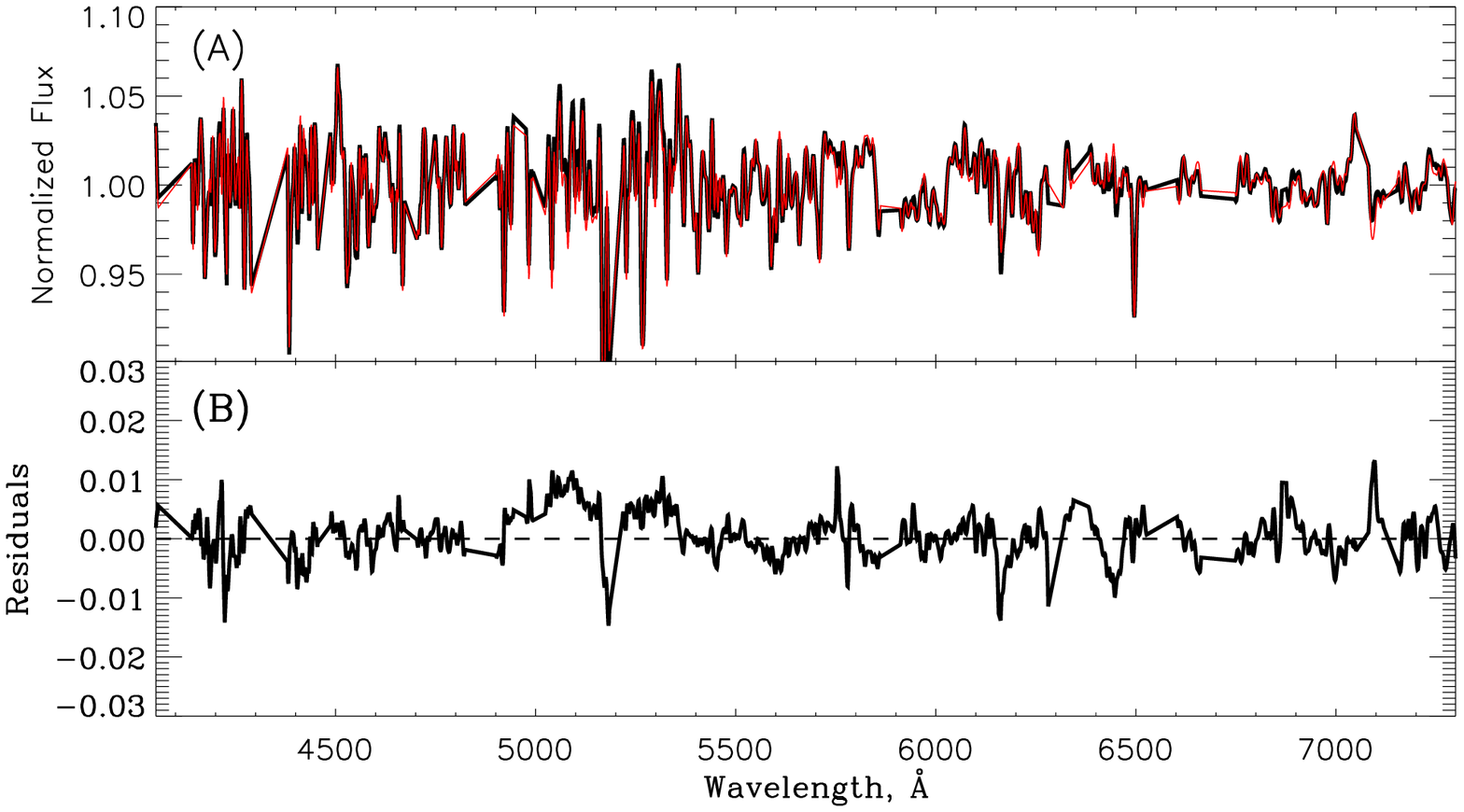}
\end{center}
\caption{(A) The black curve shows the stacked spectrum for galaxies with $10^{10} < M_\ast/M_\odot <10^{10.1}$ and the red curve is the best-fit LBSPS model. Both spectra are normalized to unity across the full wavelength range by a high-degree polynomial fit (see text for detail). (B) Residuals of the fit in (A). {Masked regions appear as straight lines.}}
\label{fig:chi2_fit}
\end{figure}

We calculate model spectra by adopting star formation and chemical evolution histories. We infer these histories using the evolutionary model presented in \citet{Zahid2012b} \citep[see also e.g.,][]{Conroy2009b, Peng2010, Papovich2011, Leitner2012}. We use these histories as inputs to the FSPS code to generate model spectra. We refer to these as the look back stellar population synthesis (LBSPS) models. The basic assumption of these models is that galaxies evolve along the average measured relations between stellar mass and SFR and stellar mass and gas-phase oxygen abundance. 

We adopt the parameterization given in \citet[parameters in Table 8]{Behroozi2013a} for the star formation rate measurements as a function of stellar mass and redshift. This parameterization is based on a large compilation of published values from the literature spanning a broad redshift range ($0<z<8$). Z14 consistently measure the $MZ_g$ relation for galaxies out to $z\sim1.6$. We adopt the relation they derive (Equations 6 - 8 in Z14). 

Figures \ref{fig:z_sfh}A and \ref{fig:z_sfh}B show the stellar mass and star formation rate as a function of cosmic time for three example model galaxies. We assume that stellar mass is instantaneously returned to the ISM and adopt a return fraction of 0.43 {which is appropriate for the Chabrier IMF} \citep{Vincenzo2016}. We evolve models to $z\sim0.08$, the median redshift of the sample. The stellar mass estimates used in Z14 and \citet{Zahid2012a} are 0.25 dex smaller than stellar masses used in this study. We add 0.25 dex to stellar masses and SFRs outputted by the evolutionary model. 

We use observations of the $MZ_g$ relation as a function of redshift to infer the chemical evolution histories of model galaxies. Z14 parameterize the relation as 
\begin{equation}
12+\mathrm{log}(O/H) = Z_{o} + \mathrm{log} \left[ 1 - \mathrm{exp} \left( -\left[ \frac{M_\ast}{M_o} \right] ^{\gamma} \right) \right].
\label{eq:mzfit}
\end{equation}
In this model, $Z_0$ is the saturation metallicity quantifying the asymptotic metallicity limit \citep{Moustakas2011, Zahid2013a}. $M_o$ is the characteristic turnover mass where the $MZ_g$ relation begins to saturate and $\gamma$ is the power-law slope of the relation at $M_\ast < M_o$. Z14 show that $Z_0$ and $\gamma$ do not evolve significantly at $z< 1.6$ and  the redshift evolution of the $MZ_g$ relation is quantified by evolution of $M_o$ which goes as $\propto (1+z)^{2.64}$. Given the uncertainty in the absolute metallicity calibration, {we adopt the $Z_0$ measured by AM13 using the direct method metallicity determination when calculating LBSPS models.} Figure \ref{fig:z_sfh}C shows the metallicity history of three example model galaxies. 

We use {offsets to} $Z_0$ in Equation \ref{eq:mzfit} as a free parameter when calculating the chemical evolution history for our evolutionary models. We do this for two reasons. First, the zero-point of the gas-phase abundance scale is uncertain \citep{Kewley2008}. Second, we want to determine the metallicity of the stellar population and, thus, we need a grid of models with a range of metallicities. For each stellar mass bin we generate a set of LBSPS models. Each set has metallicity offsets to $Z_0$ by a constant factor $\Delta Z_0$, such that $-0.75 \leq \Delta Z_0 \leq +0.75$ evenly spaced by 0.05 dex. {Because we the adopt the $Z_0$ measured by AM13, $\Delta Z_0 = 0$ corresponds to a saturation metallicity of $12 + \mathrm{log}(O/H) = 8.798$ (i.e., $[Z/Z_\odot] = 0.108$).} An example is shown in Figure \ref{fig:chi2_template}. The chemical evolution histories shown all evolve according to the Z14 relation, but the different color curves correspond to 0.05 dex offsets in $Z_0$ of Equation \ref{eq:mzfit}.

Our approach yields model spectra with different final metallicities for each stellar mass bin. Figure \ref{fig:chi2_template}B shows an example of model spectra in a small spectral window along with the stacked spectrum of galaxies with $10^{10} M_\odot <M_\ast <10^{10.1} M_\odot$. The flux variation in the LBSPS models is due solely to metallicity; the star formation history for the models are identical. We determine the best-fit metallicity by minimizing the residuals between the various models and the observed stacked spectrum. Figure \ref{fig:chi2_template}C shows the root-mean-square (RMS) difference between the observed spectrum and the models. The RMS shows a clear minimum as a function of metallicity. Here the metallicity being plotted is the final metallicity of the LBSPS model. The stacked spectrum at all stellar masses have well defined minima in the RMS similar to Figure \ref{fig:chi2_template}C. 

Figure \ref{fig:chi2_fit} shows the stacked spectrum over the full wavelength range analyzed along with the best-fit model. The data are well fit across the full wavelength range.

We fit stacked data at each stellar mass following the procedure outlined in Figure \ref{fig:chi2_template}. The RMS of the best-fit ranges between $3-6\times10^{-3}$. We quote the final metallicity of the LBSPS model as the metallicity measurement. We interpret this metallicity as the metallicity of the young stellar population. 

We estimate the systematic uncertainty in metallicity is $\sim0.1$ dex. The major sources of uncertainty are the absolute calibrations of stellar masses and SFRs. We vary these quantities by 0.25 dex and generate new LBSPS models. We find that metallicities vary systematically by $\sim0.1$ dex; the shape of the $MZ_g$ relation is robust to uncertainties in the absolute calibration of stellar masses and SFRs.

\begin{figure}
\begin{center}
\includegraphics[width = \columnwidth]{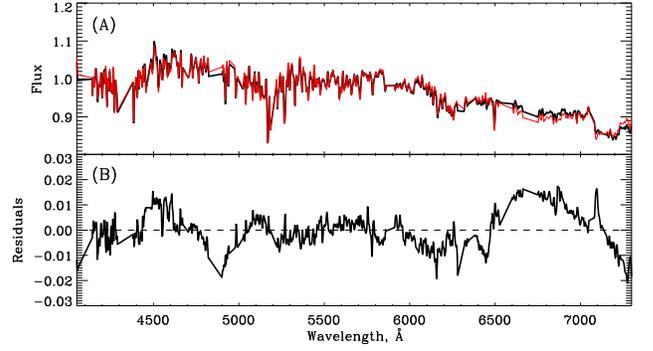}
\end{center}
\caption{(A) The black curve shows the stacked spectrum for galaxies with $10^{10} < M_\ast/M_\odot <10^{10.1}$ and the red curve is the best-fit SSB model. Both spectra are normalized to the average flux between $4400 - 4450\mathrm{\AA}$. (B) Residuals of the fit in (A). {Masked regions appear as straight lines.}}
\label{fig:chi2_fit2}
\end{figure}


\subsubsection{{Metallicity and Dust Attenuation from Sequential Single-Burst Models}}

Metallicities determined from fitting the LBSPS models provide a fiducial measurement. However, these models assume star formation and chemical evolution histories. Therefore, as an alternative approach, we also fit data with models calculated from a linear combination of sequential single burst (SSB) model spectra. These models are sampled at a range of metallicities and ages. Unlike LBSPS model fits, the star formation history and ages of galaxies are free parameters in this approach. We generate a grid of 216 single burst models sampled at 12 metallicities ranging between $-2.5 \leq \mathrm{log}(Z/Z_\odot) \leq 0.5$ spaced at the intrinsic sampling of the MIST isochrones and 18 stellar population ages logarithmically spaced between 0.001 and 12.6 Gyr. 

The shape of the stellar continuum is set by the star formation history and dust attenuation. The star formation history constrains the relative contribution of stars at different stellar masses and dust preferentially absorbs blue light making galaxies appear redder. To use this information when applying the SSB model fits, we use the shape of the continuum to additionally measure dust attenuation. This is different from the LBSPS approach described in the previous section. The relative flux of the model spectrum as a function of wavelength in this case is not normalized by fitting a high degree polynomial. The relative model flux is calculated as
\begin{equation}
M_\lambda = \frac{1}{C_\lambda(A_v)} \sum_{i=1} b_i f_{\lambda, i}.
\end{equation}
Here, the sum is over all single burst models $f_{\lambda, i}$. {The models are normalized to the mean flux between $4400 - 4450\mathrm{\AA}$; this is the same normalization applied to the observed spectra}. Each model is scaled by $b_i$ and $C_\lambda$ is the attenuation correction to the flux which we parameterize by the visual attenuation, $A_v$. We adopt the \citet{Cardelli1989} extinction law and a corresponding selective extinction ratio of $R_v=3.1$ to determine $C_\lambda$. {We use the \citet{Cardelli1989} extinction law as it is applies to both dense and diffuse regions of the interstellar medium and we are interested in consistently analyzing stellar and nebular attenuation.} The free parameters of the model are $A_v$ and $b_i$. We determine the best-fit by minimizing residuals using the MPFIT set of routines in IDL \citep{Markwardt2009}. 

Each single burst model $f_{\lambda, i}$ is sampled at a metallicity $Z_i$. We derive metallicity for each stacked spectrum from the best-fit parameters
\begin{equation}
\mathrm{log}\left[ Z(M_\ast)\right] = \sum_{i=1} b_i \mathrm{log}(Z_i).
\label{eq:z}
\end{equation}   
Here, $Z$ is the derived metallicity. 

SSB model fits also constrain ages of the stellar populations. However, we have optimized our approach to determine metallicity and thus have masked out strong Balmer absorption lines. These absorption lines are age sensitive features of the continuum but in star-forming galaxies, they are contaminated in their age sensitive line cores by strong ISM emission lines. We find our procedure is not sensitive to stellar population age and thus we do not report the age here.


Figure \ref{fig:chi2_fit2} shows a SSB model fit to a stacked spectrum. The RMS of all fits ranges between $5-10\times10^{-3}$; about twice the value of the LBSPS fits. {A mismatch between the SSB model fits and stacked spectra is seen at wavelengths $>6500 \mathrm{\AA}$ in all the fits and contributes significantly to the larger RMS of the SSB fits as compared to the LBSPS fits. Fluxes of the MILES model spectra are corrected for second-order contamination redward of $6700\mathrm{\AA}$ \citep[for details see][]{Sanchez-Blazquez2006}. One possible source for the mismatch between SSB models and observed spectra may be a small amount of residual contamination in the model spectra.}

\begin{figure}
\begin{center}
\includegraphics[width = \columnwidth]{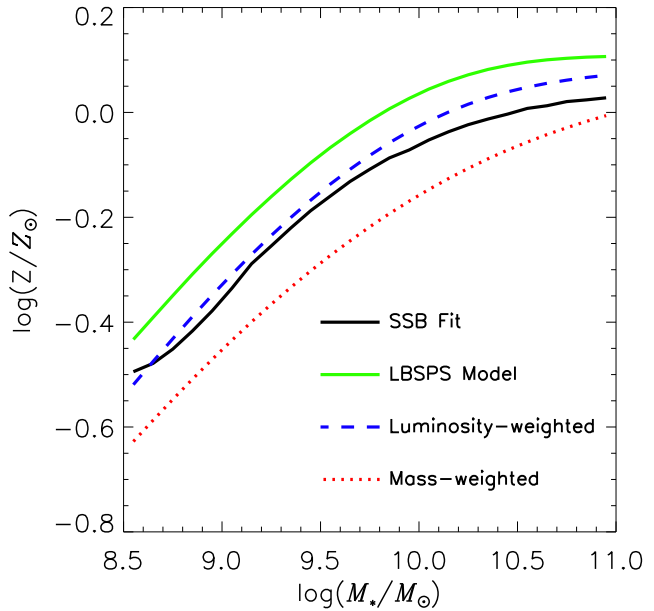}
\end{center}
\caption{Comparison of metallicities determined from chemical evolution histories used to derive LBSPS models and metallicities derived from SSB model fits. The green curve is the metallicity of the most recently formed stellar population of the LBSPS models with $\Delta Z_{o} = 0$. The blue dashed and red dotted curves are luminosity- and mass-weighted metallicities and the black curve is metallicity determined from fitting SSB models to the LBSPS models.} 
\label{fig:fit_model}
\end{figure}

\begin{figure*}
\begin{center}
\includegraphics[width = 2\columnwidth]{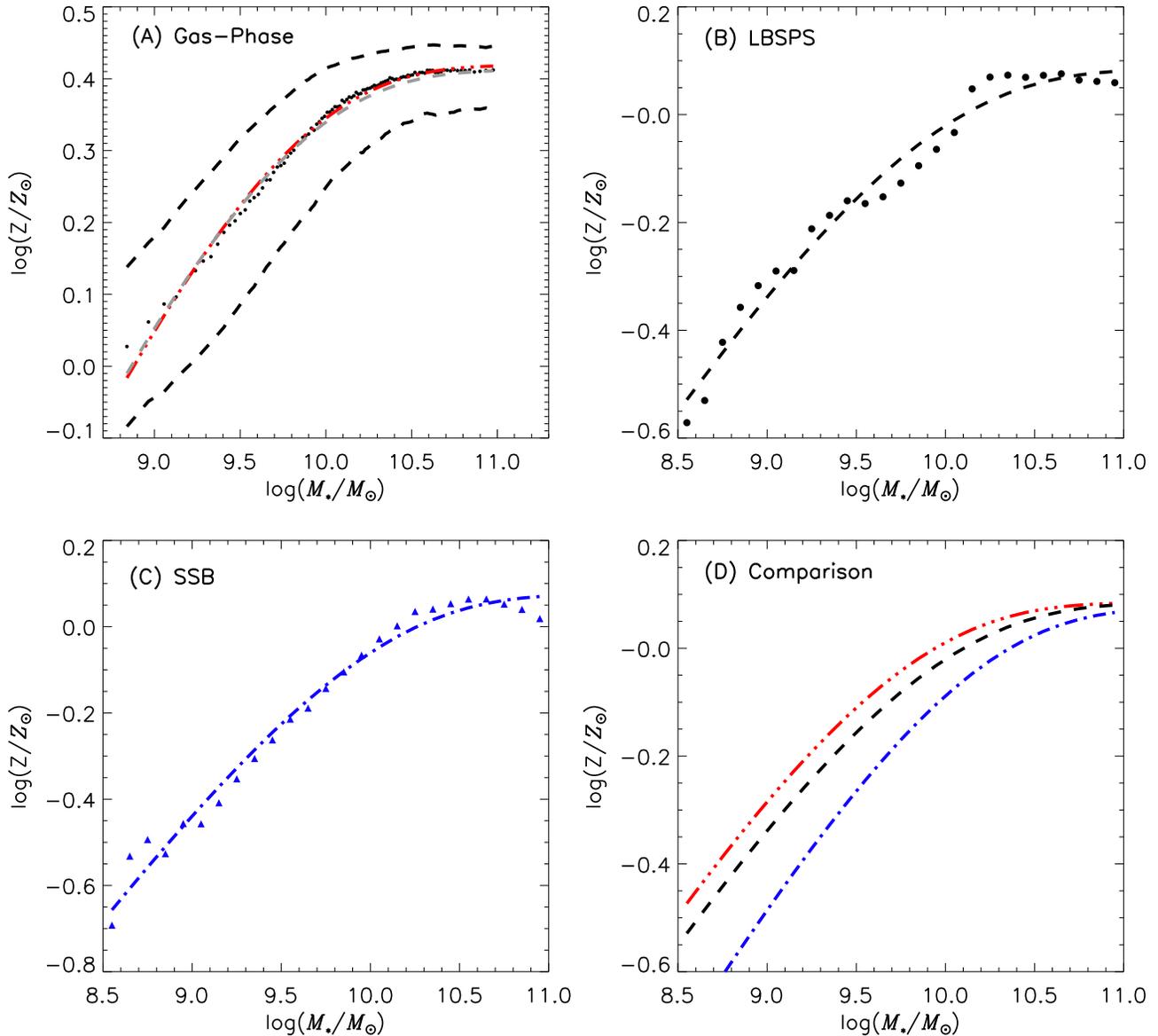}
\end{center}
\caption{(A) $MZ_g$ relation from emission line fitting. Black points are median metallicities for galaxies sorted into 100 equally populated bins of stellar mass. The dashed black line shows limits containing the central 68\% of the galaxy distribution. The triple dot dashed red curve is the best-fit of the model given in Equation \ref{eq:mzfit}. The dashed gray line is the best-fit determined by Z14 but with 0.25 dex added to the stellar mass to account for absolute offsets in stellar mass estimates. (B) $MZ_\ast$ relation determined by fitting LBSPS models. The dashed black curve is the best-fit of Equation \ref{eq:mzfit}. {The median statistical uncertainty in metallicity determined from bootstrapping is 0.0003 dex.} (C) $MZ_\ast$ relation determined by fitting SSB models. The dot dashed blue curve is the best-fit of Equation \ref{eq:mzfit}. {The median statistical uncertainty in metallicity determined from bootstrapping is 0.02 dex.} (D) Comparison of the $MZ$ relations shown in (A)-(C). Emission line metallicities are shifted by -0.33 dex to match the saturation metallicity determined from fitting LBSPS models.}
\label{fig:mz}
\end{figure*}



The metallicity we derive by fitting stacked data with SSB models represent an average global property of the stellar population. To assist in our interpretation of metallicities derived from Equations \ref{eq:z}, we fit the LBSPS models with our SSB fitting procedure. We fit the LBSPS models with stellar masses $10^{8.5} \leq M_\ast /M_\odot \leq 10^{11}$ and  $\Delta Z_{0} = 0$. We fit LBSPS model spectra with SSB spectra because the metallicity of the LBSPS model stellar population is known. Thus, we can compare the metallicity derived from the SSB fit with known quantities to understand what stellar population metallicity we are sensitive to with our SSB fitting procedure.

Figure \ref{fig:fit_model} shows the metallicity of the LBSPS stellar population compared to the metallicity derived from the SSB model fits. The green curve shows the inputted metallicity for the most recently formed stellar population contributing to the model spectra. Thus, the green curve is the metallicity of the young stellar population. The blue dashed and red dotted curves are the $V-$band luminosity- and mass-weighted stellar metallicities calculated from the star formation and chemical evolution histories used to generate the LBSPS models. The black curve is the metallicity determined by fitting SSB models to the LBSPS models. The SSB fit metallicity is most consistent with the luminosity-weighted metallicity. By comparing the SSB fit results with the luminosity-weighted metallicity in Figure \ref{fig:fit_model}, we estimate that the metallicity derived from the SSB fit systematically varies from the luminosity-weighted quantity by $\lesssim0.04$ dex.



\section{The Mass-Metallicity Relation}

\begin{deluxetable}{cccc}
\tablewidth{180pt}
\tablecaption{$MZ_\ast$ Relation and Attenuation Data}
\tablehead{ \colhead{$\mathrm{log}(M_\ast/M_\odot)$}  & \colhead{LBSPS} & \colhead{SSB} & \colhead{$A_v$} \\
                                                                                         &  $[Z/Z_\odot]$ & $[Z/Z_\odot]$  }
\startdata

 8.55 & -0.572 & -0.692 & 0.118 \\
 8.65 & -0.530 & -0.532 & 0.178 \\
 8.75 & -0.422 & -0.494 & 0.210 \\
 8.85 & -0.357 & -0.526 & 0.222 \\
 8.95 & -0.317 & -0.456 & 0.239 \\
 9.05 & -0.290 & -0.457 & 0.251 \\
 9.15 & -0.289 & -0.408 & 0.293 \\
 9.25 & -0.212 & -0.353 & 0.323 \\
 9.35 & -0.187 & -0.306 & 0.351 \\
 9.45 & -0.160 & -0.263 & 0.381 \\
 9.55 & -0.165 & -0.214 & 0.415 \\
 9.65 & -0.152 & -0.189 & 0.454 \\
 9.75 & -0.127 & -0.143 & 0.499 \\
 9.85 & -0.095 & -0.105 & 0.545 \\
 9.95 & -0.064 & -0.066 & 0.598 \\
10.05 & -0.033 & -0.028 & 0.642 \\
10.15 & 0.048 & 0.002 & 0.676 \\
10.25 & 0.070 & 0.035 & 0.713 \\
10.35 & 0.073 & 0.041 & 0.726 \\
10.45 & 0.069 & 0.053 & 0.749 \\
10.55 & 0.073 & 0.064 & 0.759 \\
10.65 & 0.076 & 0.064 & 0.772 \\
10.75 & 0.064 & 0.052 & 0.766 \\
10.85 & 0.062 & 0.040 & 0.754 \\
10.95 & 0.059 & 0.019 & 0.728 \\

\enddata
\label{tab:data1}
\end{deluxetable}

We examine the relation between stellar mass and metallicity by applying the three methods outlined in the previous section. We determine metallicities from emission lines and absorption lines.  We critically compare $MZ$ relations obtained with the three methods.

We start by analyzing the strong ISM emission lines. We measure the gas-phase metallicity for galaxies in the sample using the method described in Section 3.1. {The $MZ_g$ relation is derived by taking the median metallicity of individual galaxies in bins of stellar mass and is shown in Figure \ref{fig:mz}A.} The red curve is a fit of the observed $MZ_g$ relation with the model given in Equation \ref{eq:mzfit}. The gray dashed curve shows the $MZ_g$ relation determined by Z14 using a different sample selection criteria. The consistency of the two relations demonstrates that the $MZ_g$ relation is robust to the selection criteria. The best-fit parameters are provided in Table \ref{tab:fit}.

Next we use LBSPS models to fit the stellar metal absorption lines in the stacked spectra using the method described in section 3.2.1. Figure \ref{fig:mz}B shows the corresponding $MZ_\ast$ relation. We fit the $MZ_\ast$ relation with the model given in Equation \ref{eq:mzfit}. The best-fit is shown and the parameters are in Table \ref{tab:fit}.

\begin{deluxetable}{lccc}
\tablewidth{\columnwidth}
\tablecaption{MZ Relation Fit Parameters}
\tablehead{\colhead{Sample} &\colhead{$[Z_0/Z_\odot]$} & \colhead{$\mathrm{log}(M_o/M_\odot)$} & \colhead{$\gamma$}  }
\startdata
Gas-Phase (Z14) & 0.412$\pm$0.002 & 9.469$\pm$0.007 & 0.513$\pm$0.009\\
Gas-phase & 0.419$\pm$0.001 & 9.486$\pm$0.003 & 0.523$\pm$0.003 \\
LBSPS & 0.084 $\pm$0.019 & 9.63$\pm$0.08 & 0.51 $\pm$0.03 \\
SSB & 0.075 $\pm$0.020 & 9.79$\pm$0.08 & 0.56 $\pm$0.03 \\
\enddata
\label{tab:fit}
\end{deluxetable}

{There is a weak dependence of the $MZ_\ast$ relation we derive using LBSPS models on the parameterization we use for chemical evolution. If we input a $\gamma = 0.6$ for $MZ_g$, the best-fit $MZ_\ast$ $\gamma = 0.57$. We test more extreme values, adopting a $\gamma = 0.7$ and $\gamma = 0.3$ for the input $MZ_g$. In this case, we find that the $MZ_\ast$ relation we fit has a $\gamma = 0.62$ and $\gamma = 0.42$, respectively. These differences are not statistically significant ($<2.5 \sigma$), but they are systematic. The statistical uncertainty on the $MZ_g$ $\gamma$ measured by Z14 is $0.009$; the systematic uncertainty is also likely to be small as a consistent $\gamma$ is measured using independent data sets out to $z\sim0.8$ (see Z14). Thus, we have tested the variations in $\gamma$ using extreme values that are unlikely to be the true values. However, we note the dependency of the output parameters on the input parameters for completeness.}

{The SSB metallicities are completely independent of the LBSPS model assumptions and thus provide a consistency check on the LBSPS metallicities we derive.} We use SSB models to fit stellar metal absorption lines in stacked spectra using the method described in section 3.2.2. Figure \ref{fig:mz}C shows the corresponding $MZ_\ast$ relation. The blue curve is a fit to the $MZ_\ast$ relation parameterized by Equation \ref{eq:mzfit}. The fit parameters are in Table \ref{tab:fit}.

Figure \ref{fig:mz}D shows a comparison of the $MZ$ relations determined from the three methods. The $MZ_g$ relation is reasonably consistent with the $MZ_\ast$ relation from fitting LBSPS models when it is shifted by -0.33 dex. This shift corresponds to the difference in the $Z_0$ we measure using the two approaches. The LBSPS model results also correspond to the metallicity of the young stellar population and thus should be consistent with gas-phase metallicity measurements. Such a shift is within the range of absolute calibration uncertainties of the strong ISM emission line diagnostics \citep{Kewley2008}. For reference, the $MZ_g$ relation derived using the direct method is similarly offset by $\sim$-0.3 dex from the $MZ_g$ relation derived using the KK04 calibration (AM13). Thus, the LBSPS $MZ_\ast$ relation is consistent to within a few tenths of a dex with the direct method $MZ_g$ relation.

The $MZ_\ast$ relation derived from the two methods using stellar metal absorption lines are in reasonable agreement, though there is an offset that scales with stellar mass. Metallicity derived using the LBSPS model fit represents the metallicity of the young stellar population whereas metallicity determined from fitting SSB models is closer to a luminosity-weighted average. Metallicity increases as galaxies evolve and thus we expect luminosity weighted metallicities to be smaller than the metallicity of the young stellar population. We see this in Figure \ref{fig:fit_model}A. The difference between the LBSPS and SSB $MZ_\ast$ relations is qualitatively consistent with this conclusion. 

\section{{Stellar Mass and Dust Attenuation}}



\begin{figure}
\begin{center}
\includegraphics[width = \columnwidth]{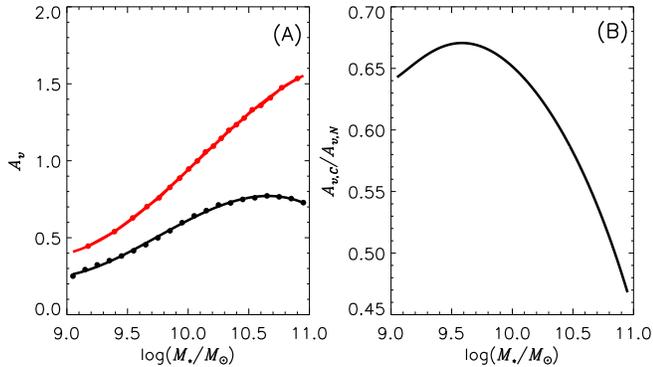}
\end{center}
\caption{(A) Attenuation determined from stellar continuum (black curve) and nebular emission lines (red curve). Stellar continuum attenuation is derived from fitting stacked spectra with SSB models and nebular attenuation is derived from the Balmer decrement. {The median statistical uncertainty in the nebular and stellar attenuation determined from bootstrapping is 0.005 and 0.01 magnitudes, respectively.} (B) Ratio of the two curves in (A).}
\label{fig:ebv_nebular_continuum}
\end{figure}

We constrain dust attenuation of the stellar population when fitting SSB models. Figure \ref{fig:ebv_nebular_continuum}A displays the relation between dust attenuation and stellar mass. The continuum attenuation is measured from fitting the stacked spectra with SSB models. Nebular emission attenuation is measured from the Balmer decrement. Both measurements show similar trends; massive galaxies are more attenuated \citep[see also,][]{Brinchmann2004, Stasinska2004, Asari2007, Garn2010b, Zahid2013a}.

Figure \ref{fig:ebv_nebular_continuum}B shows the continuum-to-nebular attenuation ratio. Nebular emission is more attenuated than the continuum and the difference scales with stellar mass; massive galaxies exhibit higher levels of nebular attenuation relative to the continuum. {Because we have used the same extinction law and selective extinction ratio to derive stellar and nebular attenuation, our comparison implicitly assumes that differences between stellar and nebular attenuation are due to variations in dust content. However, the observed difference could also result from spatial variations in the dust properties themselves in which case, a different extinction law may need to be used to derive stellar and nebular attenuation. It is beyond the scope of this work to investigate these two possibilities in detail and we interpret the results primarily as a spatial variation in dust content.}

\citet{Calzetti1994} find that attenuation determined from the Balmer line ratio is a factor of $\sim2$ larger than attenuation determined from the UV continuum. They suggest that this is due to an inhomogeneous dust distribution. Young, massive stars are typically found in dustier regions than the underlying stellar population. The canonical ratio of continuum-to-nebular attenuation is 0.44 \citep{Calzetti1997b}. However, studies suggest that this ratio may be a function of galaxy properties \citep[e.g.,][]{Wuyts2013, Koyama2015}. Based on ultraviolet to infrared observations, \citet{Koyama2015} find that for most of their sample the ratio varies between 0.44 and 1 and the ratio tends to be smaller for the most massive galaxies. Results in Figure \ref{fig:ebv_nebular_continuum} are consistent with this type of variation in the ratio. Our results pertain to the central regions of galaxies covered by the SDSS fiber. However, we have checked that there is no strong aperture bias in the attenuation we derive and the consistency with \citet{Koyama2015} suggest our results are indicative of global galaxy properties.

\section{{Stellar Mass, Metallicity, Dust Extinction and Star Formation Rate}}

\begin{figure*}
\begin{center}
\includegraphics[width = 2\columnwidth]{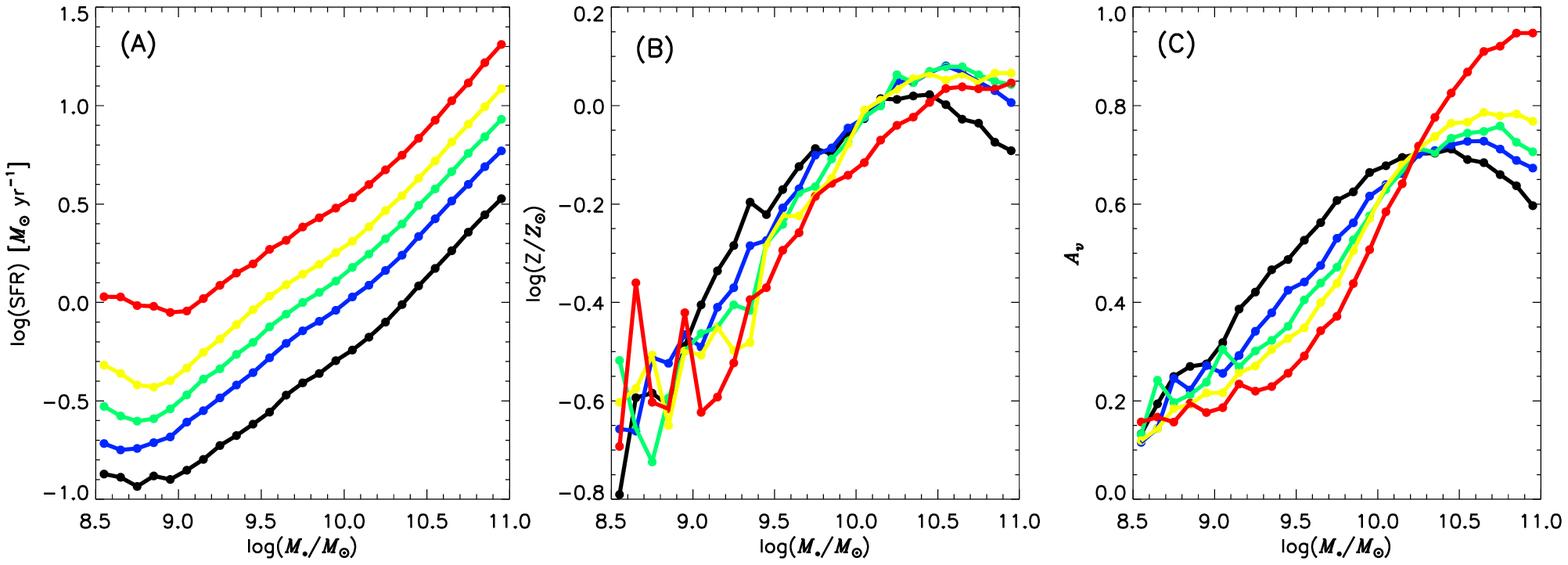}
\end{center}
\caption{(A) Median SFR corresponding to stacked spectra sorted in quintiles of SFR. (B) $MZ_\ast$ relation and (C) stellar mass - dust attenuation relation as a function of stellar mass and SFR determined from fitting SSB models. The colored curves in (B) and (C) correspond to SFRs shown in (A). {The median statistical uncertainty determined from bootstrapping in (B) and (C) is 0.05 dex and 0.02 magnitudes, respectively.}}
\label{fig:z_sfr}
\end{figure*}

We examine the relation between stellar mass, stellar metallicity, dust attenuation and SFR by analyzing stacked spectra sorted into quintiles of SFR. We fit stacked stellar absorption line spectra with SSB models to determine metallicity and dust attenuation. Figure \ref{fig:z_sfr}A shows the median SFRs of the stacked data.

Figure \ref{fig:z_sfr}B shows the $MZ_\ast$ relation for galaxies as a function of SFR. At the lowest stellar masses, the data are noisy. At intermediate stellar masses, there appears to be a weak anti-correlation between metallicity and SFR; galaxies with high SFRs have lower metallicities and vice versa. Similar trends between metallicity and SFR are reported for galaxies when metallicities are determined using strong emission lines \citep{Mannucci2010, Lara-Lopez2010, Yates2012}. 


Figure \ref{fig:z_sfr}C shows the relation between stellar mass, SFR and dust attenuation. At lower stellar masses ($<10^{10.2} M_\odot$), dust attenuation is anti-correlated with SFR; similar to the anti-correlation between metallicity and SFR. There is a sharp transition in the relation at $M_\ast \sim 10^{10.2} M_\odot$ and the trend reverses at higher stellar masses such that there is a positive correlation between dust attenuation and SFR. 

 \begin{figure}
\begin{center}
\includegraphics[width = \columnwidth]{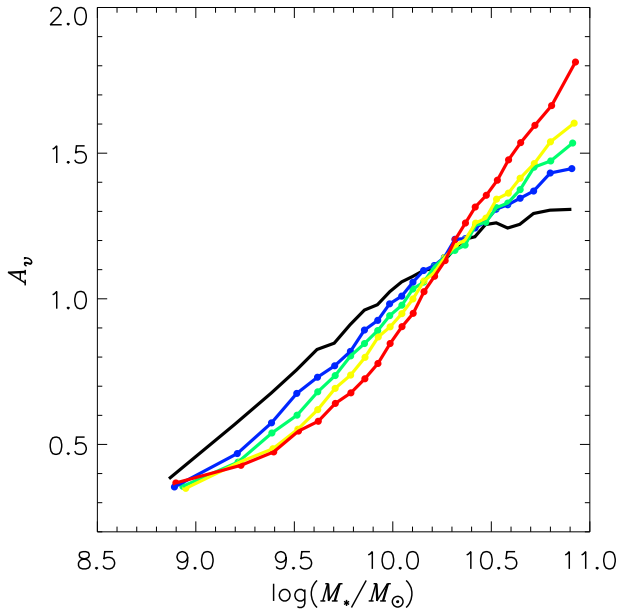}
\end{center}
\caption{Dust attenuation determined from the Balmer decrement as a function of stellar mass and SFR. Note the consistency in the trends based on emission line analysis with those based on stellar continuum analysis (see Figure \ref{fig:z_sfr}). {The median statistical uncertainty determined from bootstrapping is 0.01 magnitudes. This figure is a reproduction of Figure 2b in \citet{Zahid2013a} using the sample analyzed in this study. }}
\label{fig:mz_ebv_sfr}
\end{figure}

The trend in dust attenuation shown in Figure \ref{fig:z_sfr}C is remarkably consistent with reported trends between stellar mass, dust attenuation and SFR based on emission line analysis \citep{Zahid2013a}. We reproduce the emission line analysis result in Figure \ref{fig:mz_ebv_sfr} using our sample of SDSS galaxies.


{ \citet[Z13 hereafter]{Zahid2013c} develop a model of dust efflux which qualitatively reproduces the observed relation between stellar mass, dust attenuation and SFR.  In the Z13 model, variations in attenuation are explicitly interpreted as a consequence of the total dust content and not a change in dust properties. Dust content is taken as the integral of the rate of dust production and loss. The model is explicitly time dependent and is based on star formation and chemical evolution histories determined using a procedure similar to one described in Section 3.2.1. The simplest parameterization of the model is:
 \begin{equation}
 M_d(t) \propto \int_{t_f}^t [\dot{M}_{R}(t^\prime) - \eta \tau (t^\prime) \Psi (t^\prime - \Delta t)] dt^\prime.
\label{eq:dust}
 \end{equation}
Here, $M_d$ is the mass of dust at some time $t$ and $t_f$ is the formation time of the model galaxy. $\dot{M}_R$ is the time dependent stellar mass recycling rate, i.e. the rate at which mass in stars is recycled back into the ISM via supernova, stellar winds, etc., and is calculated from star-formation histories which are explicitly time dependent. $\eta$ is the efficiency of dust loss, $\tau$ is the dust optical depth and $\Psi$ is the SFR. Both $\Psi$ and $\tau$ are calculated using look back models and are explicitly time dependent. $\Delta t$ represents a timescale over which dust is effluxed out of the line-of-sight; $\Delta t$ characterizes the outflow velocity. }

{In the Z13 model, the dust content is set by the difference in the integrated rate of production and rate of dust loss as parameterized by Equation \ref{eq:dust}. Dust is produced from stellar material recycled back to the ISM. Thus, the first term on the right-hand-side (RHS) of Equation \ref{eq:dust} is proportional to the rate of dust production. Dust is effluxed by the sustained interaction between dust particles and the ambient radiation field. Thus, the rate depends on the instantaneous dust content and the SFR. The second term on the RHS is proportional to the rate of dust efflux. $\eta$ and $\Delta t$ are free parameters of the model and are tuned to reproduce the observed relation between stellar mass, attenuation and SFR. The primary constraints for the model are the stellar mass at which the ``twist" or ``turnover" in the relation occurs and the spread in attenuation at large stellar masses. Z13 derive a $\Delta t$ yielding an effective outflow velocity of $\sim 1$ km s$^{-1}$ and thus refer to their model as the ``slow flow" model.}

{ A key assumption of the Z13 model is that the scatter in the relation between stellar mass and SFR is temporally correlated; offsets from the mean relation are persistent in time. As a consequence, at a fixed stellar mass, galaxies with large SFRs are younger and vice versa. When $\tau$ is small (i.e. at low stellar masses) the dust efflux term in Equation \ref{eq:dust} is negligible; the dust content depends primarily on the mass of material recycled back into the ISM by stars. There is an anti-correlation between attenuation and SFR at low stellar masses because galaxies with high SFRs are younger and thus have recycled a smaller fraction of their stellar mass back into the ISM. On the other hand, at larger stellar masses, the dust content of galaxies is larger and thus the dust efflux term is not negligible. The instantaneous rate of dust efflux is proportional to the SFR and thus larger in galaxies with high SFRs. However, given a sufficiently long timescale for dust efflux, the total integrated dust loss is larger in older, low SFR galaxies. Hence, the positive correlation between SFR and attenuation. Massive galaxies with larger SFRs are younger and thus have had less time to efflux dust.}

Dust attenuation measured from stellar continuum analysis presented in this work provides independent confirmation of the results presented in \citet{Zahid2013b} and Z13. {We note the qualitative consistency between the Z13 model and results presented here. It is beyond the scope of this work to explicitly tune the model parameters to fit the observations or explore alternative models based on variations in dust properties.}

 
\section{Discussion}

The $MZ_g$ relation based on emission line analysis of star-forming galaxies in the SDSS has been extensively studied \citep[e.g.,][]{Tremonti2004, Kewley2008, Yates2012, Andrews2013}. We critically compare the $MZ_\ast$ relation derived from stellar continuum absorption line analysis with the $MZ_g$ relation. We find that after accounting for absolute uncertainties in the zero-point of the gas-phase metallicity calibration, the $MZ_\ast$ relation measured from the stellar continuum is consistent with the relation derived from emission lines. The stellar metallicities we derive are completely independent of the emission line analysis and thus the consistency in the stellar and nebular $MZ$ relations for star-forming galaxies is an important confirmation of previously reported results. 
 
 The shape of the $MZ$ relation provides important constraints for understanding the chemical evolution of galaxies. Z14 posit that the $MZ_g$ relation originates from a more fundamental relation between metallicity and stellar-to-gas-mass ratio. Their inflow model \citep[see Equation \ref{eq:mzfit} and ][]{Larson1972} suggests that chemical evolution is characterized by three distinct regimes: gas-rich, gas-poor and gas-depleted. 
 
 Less massive galaxies tend to be gas-rich and their metallicity is proportional to the stellar-to-gas-mass ratio. Thus, the Z14 model predicts that the power-law slope of the $MZ_g$ relation ($\gamma$ in Equation \ref{eq:mzfit}) should be set by the slope of the relation between gas mass and stellar mass. Based on the measured slope of the stellar-to-gas-mass relation \citep[e.g.,][]{Peeples2014}, the $MZ_g$ relation slope should be $\gamma \sim 0.5$ which is consistent with our measurements (see Table \ref{tab:fit}).
 
 In the gas-poor regime, the stellar mass exceeds the gas mass. The metallicity increases as gas is consumed and a progressively larger fraction of the ISM metals become locked up in low mass stars; the $MZ_g$ relation begins to saturate. In the gas-depleted regime, metallicity can become large because the gas content is small and thus does not dilute the metal reservoir. At some point, the mass of metals produced by massive stars equals the mass of metals locked up in low mass stars. The metallicity can not increase beyond this level and the $MZ_g$ relation saturates. Saturation of the $MZ_\ast$ relation should occur at larger stellar masses as compared to the $MZ_g$ relation. This is because the gas-phase abundances are an instantaneous measure of the metallicity whereas stellar metallicities are an integrated property. Saturation occurs only when a large enough fraction of stars form in the saturation regime. We find that the $MZ_\ast$ relation saturates at a stellar mass that is $0.2 - 0.3$ dex larger than the $MZ_g$ relation.
 
Several studies report a dependence of the $MZ_g$ relation on SFR \citep{Mannucci2010, Lara-Lopez2010, Yates2012, Bothwell2013, Salim2014}. At stellar masses below the saturation metallicity, all these studies report an anti-correlation between metallicity and SFR. A straightforward interpretation of this dependence is that higher gas fractions dilute the metal reservoir, resulting in lower metallicity. Higher gas fractions also support larger SFRs, thus the anti-correlation between metallicity and SFR. A similar, albeit weaker, anti-correlation exists between stellar metallicity and SFR at a fixed stellar mass (see Figure \ref{fig:z_sfr}A). 

Variations in gas content appear to drive the correlation between metallicity and SFR at a fixed stellar mass. Gas content variations probably result from variations in the accretion history of galaxies \citep{Dutton2010}. The timescale of the deviation of galaxy gas content from the population average must be sufficiently long such that scatter in the $MZ_g$ relation which is correlated with SFR can be measured in integrated stellar light. Detailed chemical evolution modeling which is beyond the scope of this work may provide constraints for this timescale.
 
 \begin{figure}
\begin{center}
\includegraphics[width = \columnwidth]{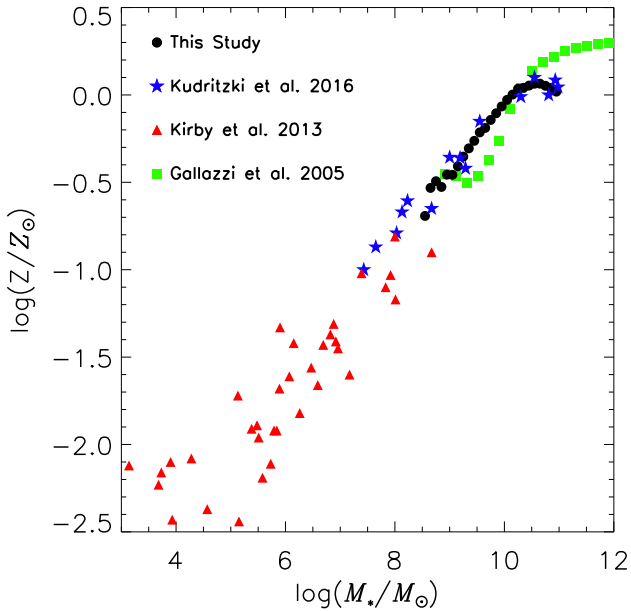}
\end{center}
\caption{Comparison of the $MZ_\ast$ relation measured by various authors. Black points are the $MZ_\ast$ relation measured by fitting SSB models to stacked spectra. Blue stars are stellar metallicities of nearby galaxies measured from spectroscopy of individual supergiant stars \citep{Kudritzki2016, Bresolin2016, Davies2017}. Red triangles are stellar metallicities measured for dwarf galaxies in the Local Group \citep{Kirby2013}. Green squares are the median $MZ_\ast$ relation from SDSS based on fitting of Lick Indices \citep{Gallazzi2005}. The \citet{Gallazzi2005} relation is derived from star-forming and quiescent galaxies. We renormalize the \citet{Gallazzi2005} metallicities to the solar metallicity adopted in this work (Z = 0.0142).}
\label{fig:comp}
\end{figure}

Figure \ref{fig:comp} shows a comparison of stellar metallicities we measure with results published in the literature. The $MZ_\ast$ relation we derive is consistent with the relation derived by \citet{Kudritzki2016} based on analysis of individual supergiant stars. With the inclusion of the \citet{Kirby2013} results, the relation appears to be continuous over $>7$ orders of magnitude in stellar mass. The \citet{Gallazzi2005} $MZ_\ast$ relation is qualitatively similar to our results but differs quantitively. The \citet{Gallazzi2005} relation is measured from Lick indices and they combine the star-forming and quiescent galaxy populations. Thus, differences may be due to systematics in the measurement technique and/or sample selection.


 \section{Conclusion}
 
 We analyze stellar absorption line continua of star-forming galaxies in SDSS to determine the relation between stellar mass, metallicity, dust attenuation and star formation rate. We stack spectra of star-forming galaxies in bins of stellar mass and fit the data with stellar population synthesis models. The continua of star-forming galaxies are remarkably consistent with model spectra calculated using empirically determined star formation and chemical evolution histories. 
 
 We also fit the data with a linear combination of single burst models sampled at a range of metallicities and ages. The star formation and chemical evolution histories of these models are not constrained. This approach yields metallicities consistent with those determined from fitting the empirically constrained models. We conclude that for star-forming galaxies the relation between stellar mass and stellar metallicity is consistent with the relation between stellar mass and gas-phase metallicity once we account for systematic uncertainties in the absolute calibration of the gas-phase metallicity. This consistency is expected from simple models of galactic chemical evolution. 
 
 We analyze the stellar metallicity and continuum dust attenuation of galaxies as a function of their star formation rate. Our results based on analyzing the continuum are consistent with previously reported trends based on emission line analysis, thus providing an independent confirmation of those results. At a fixed stellar mass, the metallicity is anti-correlated with star formation rate which is consistent with the results of e.g., \citet{Mannucci2010} and is expected if there is a universal relation between metallicity and stellar-to-gas mass ratio as suggested by \citet{Zahid2014a}. The dependence of the dust attenuation of the continuum on star formation rate is also consistent with analysis based on the Balmer decrement \citep{Zahid2013a}. 
 
 
Our analysis provides a framework for understanding the relation between gas-phase and stellar metallicities of star-forming galaxies. The results pave the way for future efforts jointly analyzing the star-forming and quiescent galaxy population. A consistent set of models and methods applied to star-forming and quiescent galaxies allows for an exploration of the means by which star-forming galaxies deplete their gas supply eventually shutting down star formation. A comprehensive study of the gas-phase and stellar metallicities of star-forming galaxies and the stellar metallicities of quiescent galaxies \citep[see e.g.,][]{Conroy2014} is already possible. The SDSS provides a wealth of spectroscopic data and the stellar population models  for star-forming galaxies are now validated. We will apply the models developed in this work to the full SDSS population to investigate how star-forming galaxies evolve to the red sequence.

\begin{center}
\begin{longtable}{cccc}
\endhead
\caption{ \\ $MZ_\ast$-SFR Relation Data} \\

\hline \hline \\

\colhead{$\mathrm{log}(M_\ast/M_\odot)$} & \colhead{SFR}& \colhead{$[Z/Z_\odot]$} & \colhead{$A_v$} \\
                                                                                         & $M_\odot$ yr$^{-1}$ & &  \\
 
 \\                                                                                        
\hline \\
                                                                                         
8.55 & -0.87 & -0.790 & 0.133 \\
 8.55 & -0.72 & -0.657 & 0.116 \\
 8.55 & -0.53 & -0.518 & 0.132 \\
 8.55 & -0.32 & -0.603 & 0.119 \\
 8.55 & 0.03 & -0.693 & 0.157 \\
 8.65 & -0.89 & -0.594 & 0.194 \\
 8.65 & -0.75 & -0.661 & 0.144 \\
 8.65 & -0.58 & -0.656 & 0.242 \\
 8.65 & -0.36 & -0.575 & 0.143 \\
 8.65 & 0.03 & -0.360 & 0.167 \\
 8.75 & -0.94 & -0.584 & 0.249 \\
 8.75 & -0.74 & -0.512 & 0.245 \\
 8.75 & -0.60 & -0.724 & 0.197 \\
 8.75 & -0.42 & -0.507 & 0.184 \\
 8.75 & -0.02 & -0.603 & 0.157 \\
 8.85 & -0.88 & -0.609 & 0.270 \\
 8.85 & -0.71 & -0.524 & 0.222 \\
 8.85 & -0.59 & -0.594 & 0.212 \\
 8.85 & -0.43 & -0.650 & 0.193 \\
 8.85 & -0.02 & -0.616 & 0.195 \\
 8.95 & -0.90 & -0.489 & 0.275 \\
 8.95 & -0.68 & -0.465 & 0.272 \\
 8.95 & -0.54 & -0.499 & 0.238 \\
 8.95 & -0.40 & -0.498 & 0.216 \\
 8.95 & -0.05 & -0.421 & 0.176 \\
 9.05 & -0.85 & -0.405 & 0.318 \\
 9.05 & -0.61 & -0.491 & 0.256 \\
 9.05 & -0.47 & -0.463 & 0.304 \\
 9.05 & -0.33 & -0.507 & 0.216 \\
 9.05 & -0.04 & -0.623 & 0.186 \\
 9.15 & -0.80 & -0.336 & 0.386 \\
 9.15 & -0.55 & -0.410 & 0.292 \\
 9.15 & -0.39 & -0.450 & 0.269 \\
 9.15 & -0.25 & -0.451 & 0.257 \\
 9.15 & 0.02 & -0.592 & 0.234 \\
 9.25 & -0.73 & -0.284 & 0.421 \\
 9.25 & -0.49 & -0.370 & 0.341 \\
 9.25 & -0.34 & -0.405 & 0.301 \\
 9.25 & -0.19 & -0.497 & 0.271 \\
 9.25 & 0.09 & -0.523 & 0.220 \\
 9.35 & -0.68 & -0.196 & 0.466 \\
 9.35 & -0.42 & -0.285 & 0.379 \\
 9.35 & -0.26 & -0.416 & 0.323 \\
 9.35 & -0.11 & -0.481 & 0.304 \\
 9.35 & 0.15 & -0.394 & 0.229 \\
 9.45 & -0.62 & -0.221 & 0.487 \\
 9.45 & -0.36 & -0.274 & 0.425 \\
 9.45 & -0.20 & -0.280 & 0.351 \\
 9.45 & -0.04 & -0.281 & 0.327 \\
 9.45 & 0.20 & -0.370 & 0.256 \\
 9.55 & -0.56 & -0.171 & 0.526 \\
 9.55 & -0.28 & -0.208 & 0.442 \\
 9.55 & -0.12 & -0.242 & 0.405 \\
 9.55 & 0.03 & -0.225 & 0.349 \\
 9.55 & 0.27 & -0.294 & 0.291 \\
 9.65 & -0.47 & -0.124 & 0.562 \\
 9.65 & -0.21 & -0.169 & 0.475 \\
 9.65 & -0.06 & -0.177 & 0.439 \\
 9.65 & 0.09 & -0.224 & 0.400 \\
 9.65 & 0.32 & -0.258 & 0.342 \\
 9.75 & -0.41 & -0.087 & 0.607 \\
 9.75 & -0.14 & -0.101 & 0.530 \\
 9.75 & 0.000 & -0.165 & 0.472 \\
 9.75 & 0.14 & -0.181 & 0.439 \\
 9.75 & 0.38 & -0.184 & 0.372 \\
 9.85 & -0.36 & -0.098 & 0.625 \\
 9.85 & -0.10 & -0.086 & 0.562 \\
 9.85 & 0.05 & -0.108 & 0.527 \\
 9.85 & 0.19 & -0.148 & 0.505 \\
 9.85 & 0.43 & -0.158 & 0.438 \\
 9.95 & -0.30 & -0.054 & 0.664 \\
 9.95 & -0.04 & -0.045 & 0.616 \\
 9.95 & 0.11 & -0.073 & 0.576 \\
 9.95 & 0.25 & -0.078 & 0.571 \\
 9.95 & 0.48 & -0.141 & 0.507 \\
10.05 & -0.24 & -0.026 & 0.678 \\
10.05 & 0.03 & -0.025 & 0.640 \\
10.05 & 0.18 & -0.023 & 0.629 \\
10.05 & 0.31 & -0.009 & 0.636 \\
10.05 & 0.53 & -0.116 & 0.584 \\
10.15 & -0.18 & 0.014 & 0.695 \\
10.15 & 0.09 & 0.004 & 0.662 \\
10.15 & 0.25 & -0.001 & 0.666 \\
10.15 & 0.38 & 0.011 & 0.680 \\
10.15 & 0.60 & -0.070 & 0.641 \\
10.25 & -0.10 & 0.013 & 0.702 \\
10.25 & 0.16 & 0.049 & 0.701 \\
10.25 & 0.32 & 0.063 & 0.713 \\
10.25 & 0.47 & 0.033 & 0.712 \\
10.25 & 0.67 & -0.040 & 0.718 \\
10.35 & -0.01 & 0.020 & 0.703 \\
10.35 & 0.24 & 0.047 & 0.709 \\
10.35 & 0.40 & 0.047 & 0.705 \\
10.35 & 0.54 & 0.055 & 0.737 \\
10.35 & 0.75 & -0.023 & 0.776 \\
10.45 & 0.08 & 0.022 & 0.712 \\
10.45 & 0.34 & 0.067 & 0.720 \\
10.45 & 0.49 & 0.070 & 0.734 \\
10.45 & 0.63 & 0.065 & 0.764 \\
10.45 & 0.83 & 0.006 & 0.826 \\
10.55 & 0.17 & 0.002 & 0.690 \\
10.55 & 0.42 & 0.081 & 0.728 \\
10.55 & 0.58 & 0.079 & 0.744 \\
10.55 & 0.72 & 0.052 & 0.766 \\
10.55 & 0.93 & 0.035 & 0.868 \\
10.65 & 0.26 & -0.028 & 0.684 \\
10.65 & 0.52 & 0.068 & 0.728 \\
10.65 & 0.66 & 0.079 & 0.748 \\
10.65 & 0.82 & 0.064 & 0.786 \\
10.65 & 1.02 & 0.038 & 0.910 \\
10.75 & 0.36 & -0.036 & 0.660 \\
10.75 & 0.60 & 0.049 & 0.712 \\
10.75 & 0.76 & 0.063 & 0.759 \\
10.75 & 0.91 & 0.047 & 0.779 \\
10.75 & 1.12 & 0.034 & 0.921 \\
10.85 & 0.45 & -0.074 & 0.637 \\
10.85 & 0.69 & 0.030 & 0.689 \\
10.85 & 0.84 & 0.049 & 0.726 \\
10.85 & 0.99 & 0.066 & 0.783 \\
10.85 & 1.22 & 0.034 & 0.947 \\
10.95 & 0.53 & -0.092 & 0.597 \\
10.95 & 0.77 & 0.006 & 0.673 \\
10.95 & 0.93 & 0.043 & 0.706 \\
10.95 & 1.09 & 0.066 & 0.768 \\
10.95 & 1.31 & 0.046 & 0.948 \\ \\

\hline 

\label{tab:data4}
\end{longtable}
\end{center}

\acknowledgements

HJZ gratefully acknowledges the generous support of the Clay Postdoctoral Fellowship. CC acknowledges support from the Packard Foundation. We thank the anonymous referee for carefully reading the manuscript and providing thoughtful and constructive suggestions that have improved the manuscript. We thank Claus Leitherer for useful discussion and calculations, Ben Johnson and Jason Wolfe for technical assistance and Margaret Geller, Evan Kirby and Rob Yates for helpful comments improving the clarity of the manuscript. This research was supported by the Munich Institute for Astro- and Particle Physics (MIAPP) of the DFG cluster of excellence "Origin and Structure of the Universe". 

Funding for SDSS-III has been provided by the Alfred P. Sloan Foundation, the Participating Institutions, the National Science Foundation, and the U.S. Department of Energy Office of Science. The SDSS-III web site is http://www.sdss3.org/. SDSS-III is managed by the Astrophysical Research Consortium for the Participating Institutions of the SDSS-III Collaboration including the University of Arizona, the Brazilian Participation Group, Brookhaven National Laboratory, University of Cambridge, Carnegie Mellon University, University of Florida, the French Participation Group, the German Participation Group, Harvard University, the Instituto de Astrofisica de Canarias, the Michigan State/Notre Dame/JINA Participation Group, Johns Hopkins University, Lawrence Berkeley National Laboratory, Max Planck Institute for Astrophysics, Max Planck Institute for Extraterrestrial Physics, New Mexico State University, New York University, Ohio State University, Pennsylvania State University, University of Portsmouth, Princeton University, the Spanish Participation Group, University of Tokyo, University of Utah, Vanderbilt University, University of Virginia, University of Washington, and Yale University.

\bibliographystyle{aasjournal}
\bibliography{/Users/jabran/Documents/latex/metallicity}

 \end{document}